\documentclass{XrU2005}
\usepackage{graphicx}

\title{Status of the XMM-Newton cross-calibration with SASv6.5.0}
\author{M. Stuhlinger}
\author{B. Altieri}
\author{M.P. Esquej}
\author{M.G.F. Kirsch}
\author{L. Metcalfe}
\author{A.M.T. Pollock}
\author{R.D. Saxton}
\author{M.J.S. Smith}
\author{A. Talavera}
\affil{ESAC, P.O. Box 50727, 28080 Madrid, Spain}
\author{S. Sembay}
\author{A.M. Read}
\author{D. Baskill}
\affil{University of Leicester, University Road, Leicester LE1 7RH, United Kingdom}
\author{F. Haberl}
\author{M. Freyberg}
\author{K. Dennerl}
\affil{MPE, Giessenbachstr.1, 85748 Garching, Germany}
\author{J. Kaastra}
\author{J.W. den Herder}
\author{C. de Vries}
\author{J. Vink}
\author{J. de Plaa}
\affil{SRON, Sorbonnelaan 2, 3584 CA Utrecht, The Netherlands}

\begin{document}

\keywords{XMM-Newton, SASv6.5, calibration}

\maketitle

\begin{abstract}
Further achievements of the XMM-Newton cross-calibration --- XMM internal
as well as with other X-ray missions --- are presented. We explain the
major changes in the new version SASv6.5 of the XMM-Newton science
analysis system. The current status of the cross-calibration of the
three EPIC cameras is shown. Using a large sample of blazars, the pn
energy redistribution at low energy could be further calibrated,
correcting the overestimation of fluxes in the lowest energy
regime. In the central CCDs of the MOSs, patches were identified at
the bore-sight positions, leading to an  underestimation of the low
energy fluxes. The further improvement in the 
understanding of the cameras resulted in a good agreement of the EPIC
instruments down to lowest energies. The latest release of the SAS software
package already includes corrections for both effects as shown in
several examples of different types of sources.
Finally the XMM internal cross-calibration is completed by the
presentation of the current cross-calibration status between
EPIC and RGS instruments.
Major efforts have been made in cross-calibrations with other X-ray
missions, most importantly with Chandra, of course, but also with
currently observing satellites like Swift.
\end{abstract}

\section{Major changes from SASv6.1 to SASv6.5}

\subsection{MOS responses now time and spatial dependent}
The major change from SASv6.1 to SASv6.5 is the new response
generation in \emph{rmfgen} for the EPIC MOS instruments. Using
SASv6.1 it has been 
found that, starting at about rev.~380 and increasing mission
duration, pn and MOS measurements diverge at low energies for most
observations. Using supernova remnants, a spatial dependency of this
low energy discrepancy was discovered. Further investigations reveal a
change in the MOS distribution behaviour at and close to the nominal
bore-sight positions (Fig.~\ref{fig:mospatch}), whereas the
redistribution outside these regions 
does not show any change. These spatial dependency indicate that
X-ray and/or focused particle radiation changed the physical performance
of the CCDs.

\begin{small}
\begin{figure}[h]
\centering
\includegraphics[width=0.9\linewidth]{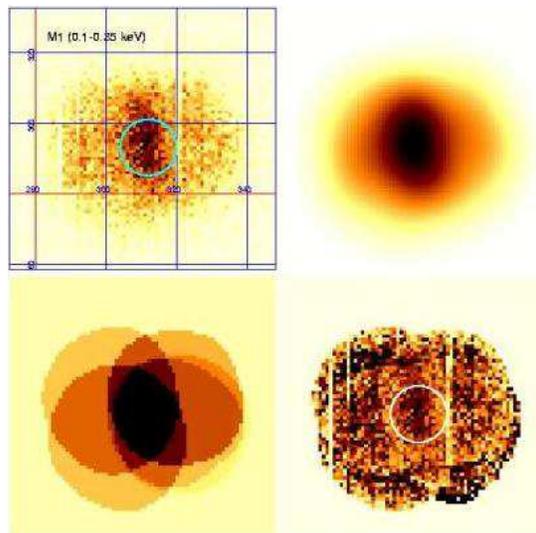}
\caption{Visualisations of the MOS patch: the chip region around the
  bore-sight positions shows time dependent redistribution behaviour.
  The new MOS response generation takes into account the time and spatial
  dependency of these central chip regions.
\label{fig:mospatch}}
\end{figure}
\end{small}

The new MOS response generation takes into account three time
dependent regions and calculates the response according to the
selected source region. Using SASv6.5 the previously found low energy
discrepancy, evolving with time, is corrected. Examples are provided
in sect.~\ref{sect:epochs}. A more detailed
description is provided in the conference contribution of A.M.~Read et
al.: 'Patching' EPIC-MOS: Temporal and Spatial Dependency of the
Detector Response.

Together with the SASv6.5 and its new MOS response generation, a new
set of 18 CCF-files has been published, for nine time epochs now
modelling the spatial dependency by three regions for the central CCDs. 
{\bf Warning: Using SASv6.1 together with the new set of CCF-files,
    \emph{rmfgen} could create responses with strange features.}

\subsection{Improvement of arfgen}
Since revolution 961 a new hot column has appeared on MOS1 CCD1 due
to an impact of a micrometeorite dust particle on MOS1 CCD1. This new
defect is leaking into the whole column. As a consequence, the offset
of this column is raised by about 20 ADUs, therefore generating a lot of
noise events at low energy above the low energy threshold, and the
whole column is identified as bad by embadpixfind and masked out in
the calibrated event list. As this column passes a few pixels from
the nominal target position on CCD1, a significant fraction of the
on-axis source PSF is affected. If the selection \#XMMEA\_EM to the
MOS event list is applied to generate spectra of on-axis sources, only
the bad column is marked bad but not the adjacent columns. This
missing column is taken into account by \emph{arfgen} in the
computation of the effective area. 

\begin{small}
\begin{figure}[h]
\centering
\includegraphics[width=0.8\linewidth]{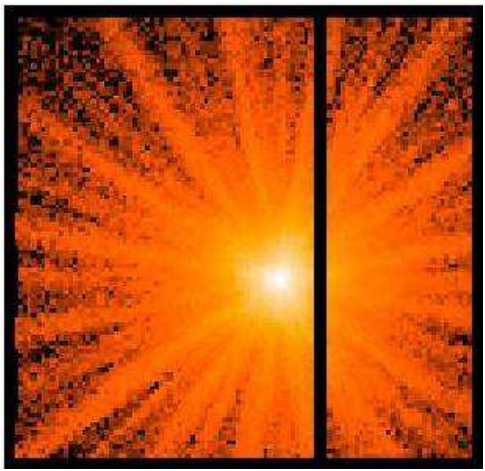}
\caption{MOS1 exposure affected by the new hot column
  after the possible micrometeorite impact of rev.~961. Using FLAG==0,
  also the two 
  adjacent columns are masked. The new \emph{arfgen} version of
  SASv6.5 now takes into account these adjacent missing columns in the
  computation of the effective area.  
\label{fig:arfgen}}
\end{figure}
\end{small}

If the more conservative selection flag FLAG==0 is used for the
analysis, also the two adjacent columns are masked, therefore 3
columns are removed, causing the loss of up to 10-15\% of the flux of
an on-axis source. The  SASv6.1 did not take into account  
these adjacent missing columns in the computation of the effective
area. Therefore the absolute flux/normalisation of a source was too
low. The new \emph{arfgen} version of SASv6.5 now takes into account
these adjacent missing columns.

\subsection{Improvement of embadpixfind}
In SASv6.1, the two MOS pipeline tasks \emph{emproc} and
\emph{emchain} used different routines to search for bad pixels. The
first used the general task \emph{badpixfind}, whereas the latter used
the more advanced, to the MOS data reduction adapted task
\emph{embadpixfind}. Thus, the resulting MOS eventlists of both
pipelines could differ distinctly. In SASv6.5, both pipeline tasks are
using the \emph{embadpixfind} routine to detect bad pixels and the
resulting event lists of both pipeline tasks are completely equivalent
now.  

\begin{small}
\begin{figure}[h]
\centering
\includegraphics[width=0.8\linewidth]{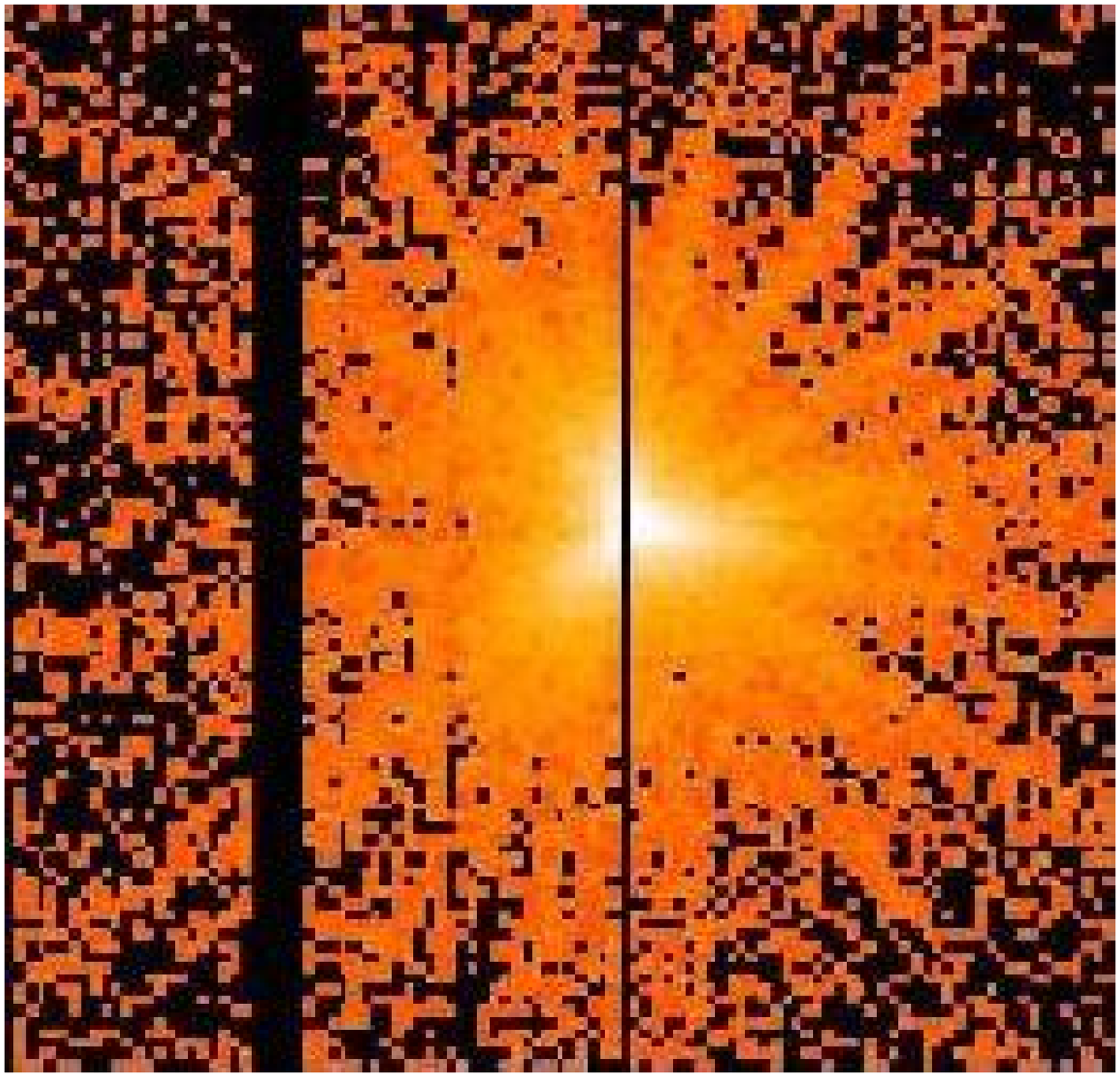}
\includegraphics[width=0.8\linewidth]{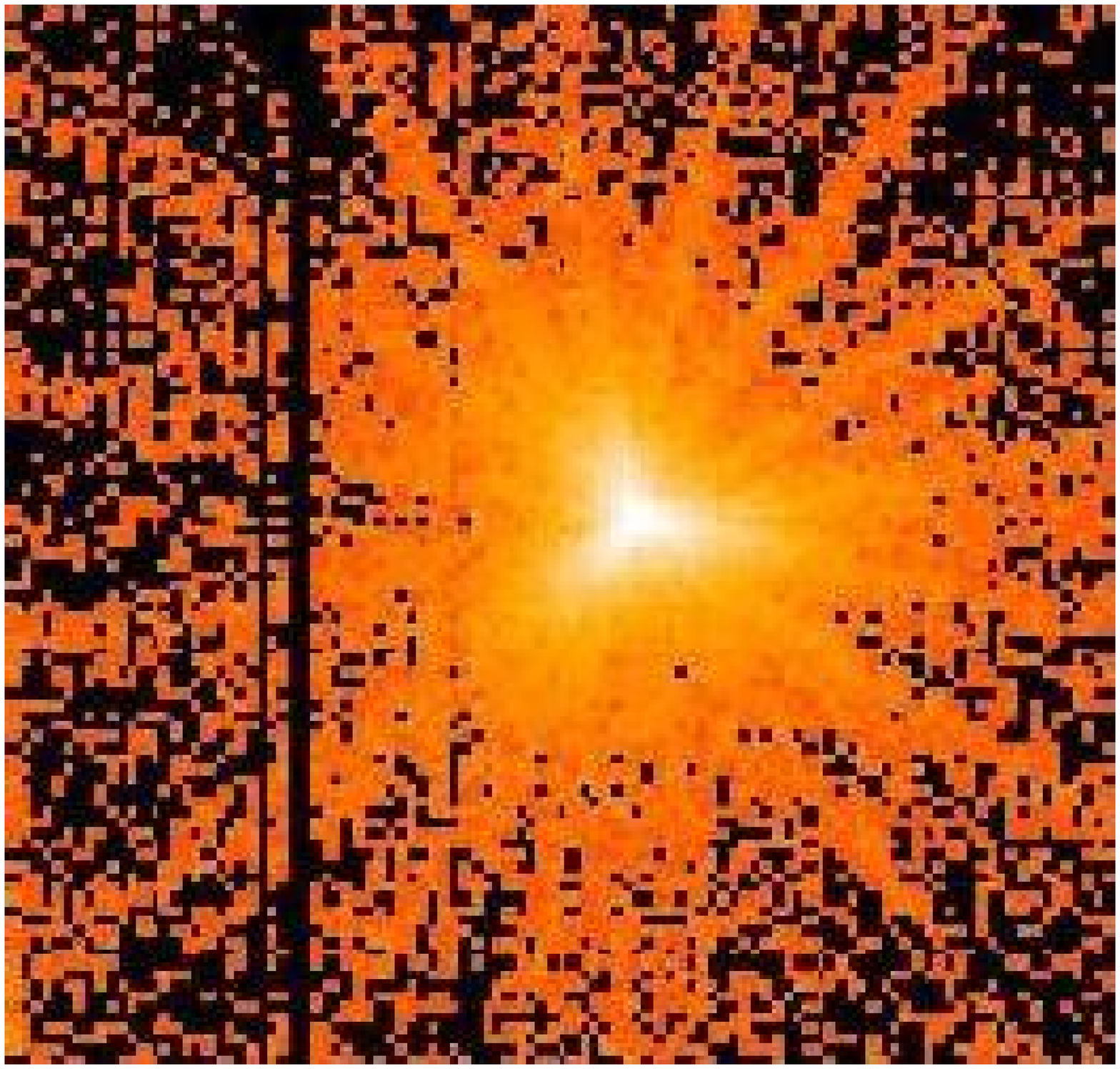}
\caption{Example for the new rejection algorithm used by
  \emph{embadpixfind}. Top: previous algorithm used in SASv6.1 could
  erroneously mark columns as bad. Bottom: SASv6.5 image of the same
  MOS2 exposure.
\label{fig:embadpixfind}}
\end{figure}
\end{small}

Due to offset variations from column to column, the previous
version of \emph{embadpixfind}, in rare cases, could remove single
columns by flagging them bad erroneously
(Fig.~\ref{fig:embadpixfind}, top). In specific observations, the 
central pixel of the PSF could be marked as a bad pixel and removed from the
event list. A new rejection algorithm prevents erroneous
identifications of bad columns and pixels
(Fig.~\ref{fig:embadpixfind}, bottom).

\section{Effect of new pn redistribution: Low energy improvement}
X-ray spectra of blazars are expected to show a featureless
continuum. Spectral fits of a set of blazars show common systematic s-shape pn
residuals at the low energy end of the pn best fits
(Fig.~\ref{fig:s-shape}). Using this set of blazars, the pn
redistribution has been optimised and already 
published via a CCF-file in 2005 May.
\begin{small}
\begin{figure}[h]
\centering
\includegraphics[width=0.9\linewidth]{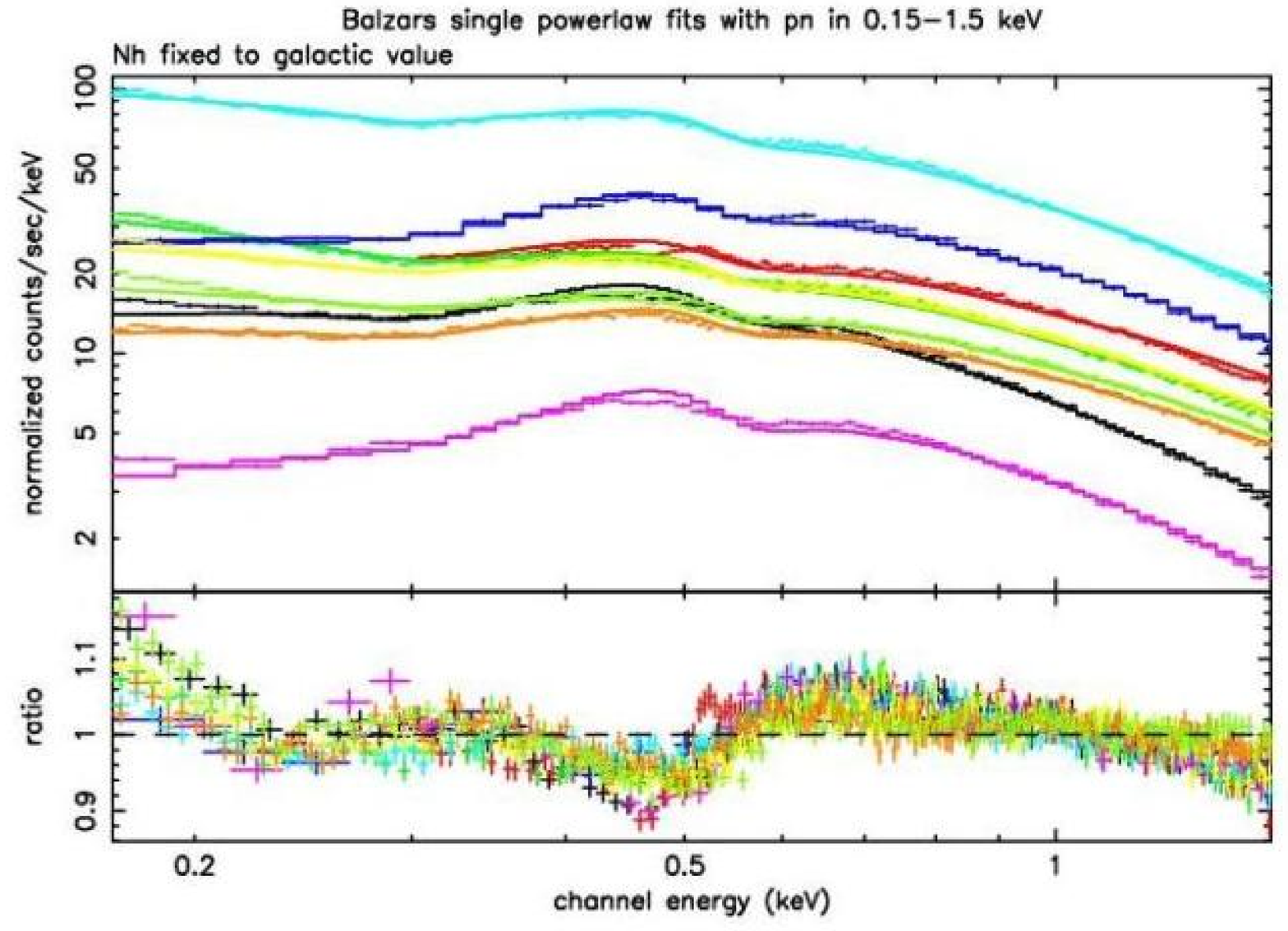}
\includegraphics[width=0.9\linewidth]{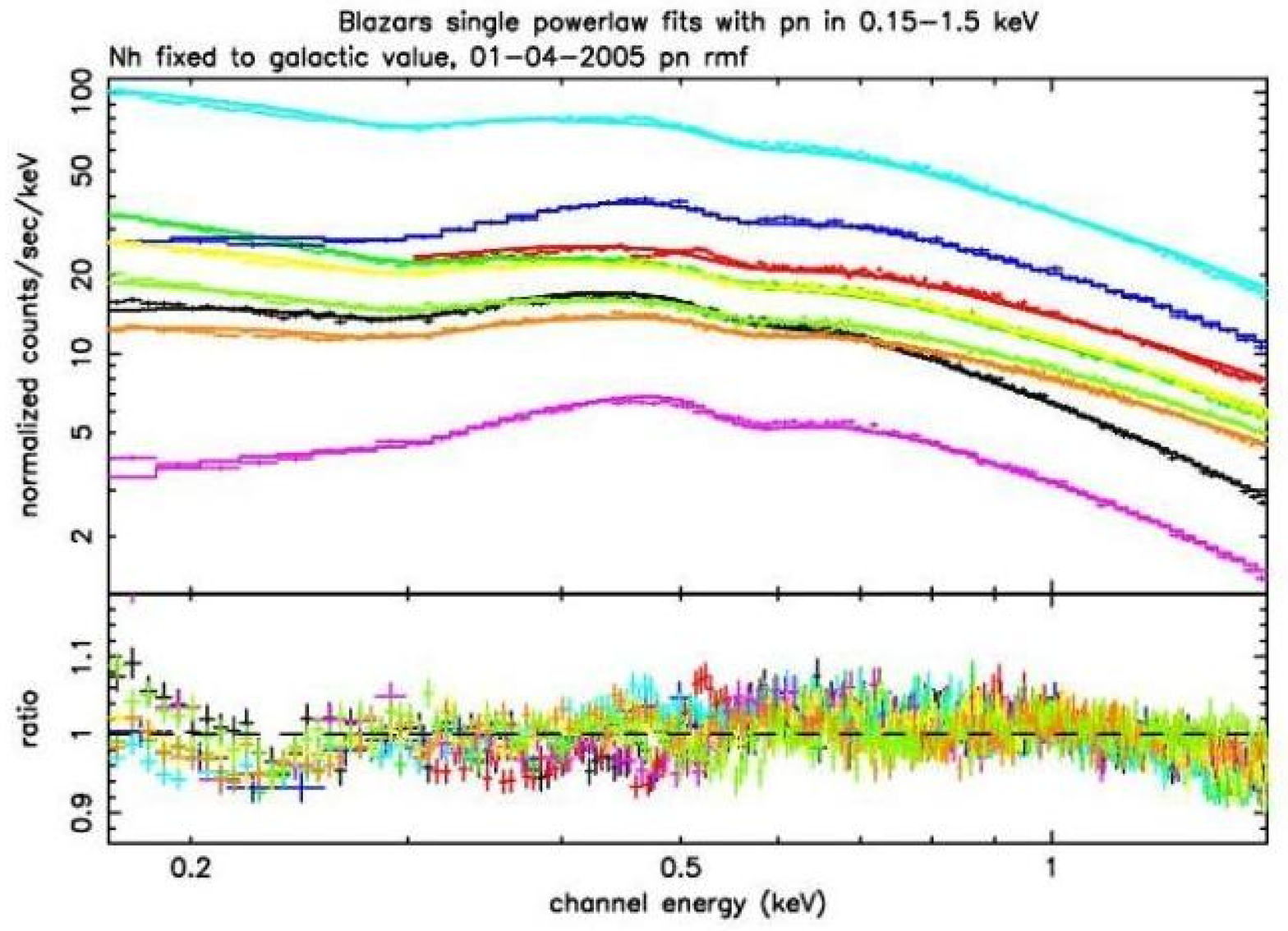}
\caption{pn best fits for a set of blazars. Top: Common s-shape
  residuals using old pn redistribution CCFs. Bottom: Flat residuals
  using new pn redistribution CCFs (published 2005 May).
\label{fig:s-shape}}
\end{figure}
\end{small}

\section{Examples for different epochs}
\label{sect:epochs}
The effect of the new MOS response generation with its time and now
spatial dependency is presented using two sources, the quasar 3C~273
as a continuum source example, and 1ES0102-7219 as a coronal source
example. 

All 3C~273 observations presented in Fig.~\ref{fig:3c273epochs} 
were performed in the EPIC small
window modes and the medium filters. As fit model, a double power law
model with galactic absorption is used. The 3C~273 series show a good
agreement of all EPIC instruments at all epochs. The time dependent low
energy discrepancy, increasing with time, is solved by the new MOS
responses. At low energies below 0.8~keV, the RGS and EPIC still
disagree. The spectral summary of the 3C~273 series is:
\begin{itemize}
\item pn bump up to 10\% between 0.4-0.5~keV
\item MOS2 bump up to 20\% at 0.2-0.4~keV
\item Above about 5~keV, pn is lower than MOS by ~10\%
\item RGS decrease by 10-20\% at lowest energies during the mission.
\end{itemize}

The low energy (0.4-0.8~keV) flux stability using 3C~273 is presented in
Fig.~\ref{fig:3c273ratios} with pn flux refered to one. The MOS to pn
ratios decrease by less than 5\% over the mission, the RGS ratios
decrease by 10-20\%.

\begin{small}
\begin{figure}[h]
\centering
\includegraphics[width=0.6\linewidth,angle=90]{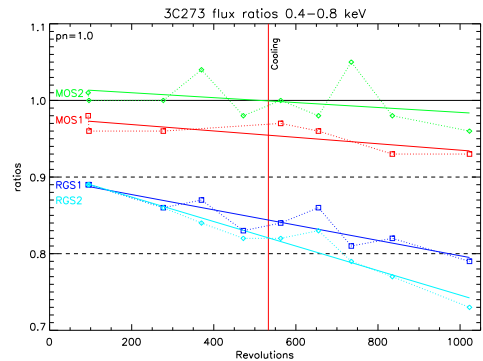}
\caption{Low energy (0.4-0.8~keV) flux stability using a series of 3C~273
  observations. There are only small changes in
  the MOS to pn ratios,  but still larger changes in the RGS/pn ratios.
\label{fig:3c273ratios}}
\end{figure}
\end{small}

All observations of 1ES0102-7219 presented in Fig.~\ref{fig:1es0102epochs}
were performed in small window mode
for pn and the large window modes for MOS. In rev.~375, 521 and 981
the thin filters were used, in rev.~888 the thick filter was used for
all EPICs. For pn small window mode, the background correction is
difficult due to the small size of the CCD window, whereas for the MOS
the background could be taken from the outer CCDs.
The fit model includes 40 lines plus absorbed bremsstrahlung. The line
energies were fixed to laboratory values and the widths are determined
by RGS. 

The 1ES0102-7219 series show that the pn response underestimates
redistribution, most evident the O-lines. Above about 0.6~keV, the
agreement between RGS and EPICs is good. For later epochs, the
decrease of the RGS low energy flux become evident again, hence the
combined fits are unreliable at low energies. In rev.~888, the thick
filter measurement, large pn-MOS discrepancies are present below
0.5~keV. 

\begin{small}
\begin{figure}
\centering
\includegraphics[width=0.8\linewidth]{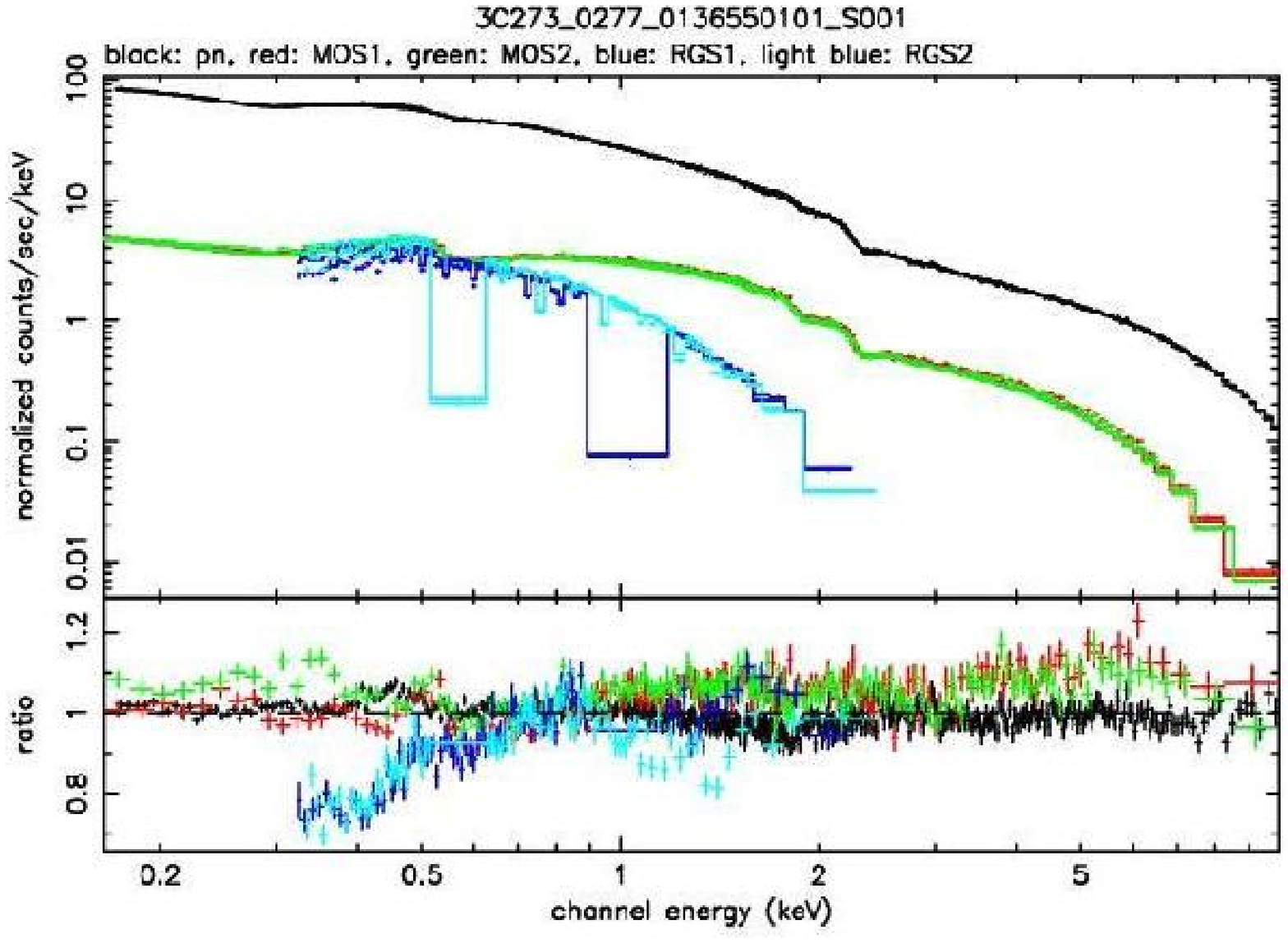}
\includegraphics[width=0.8\linewidth]{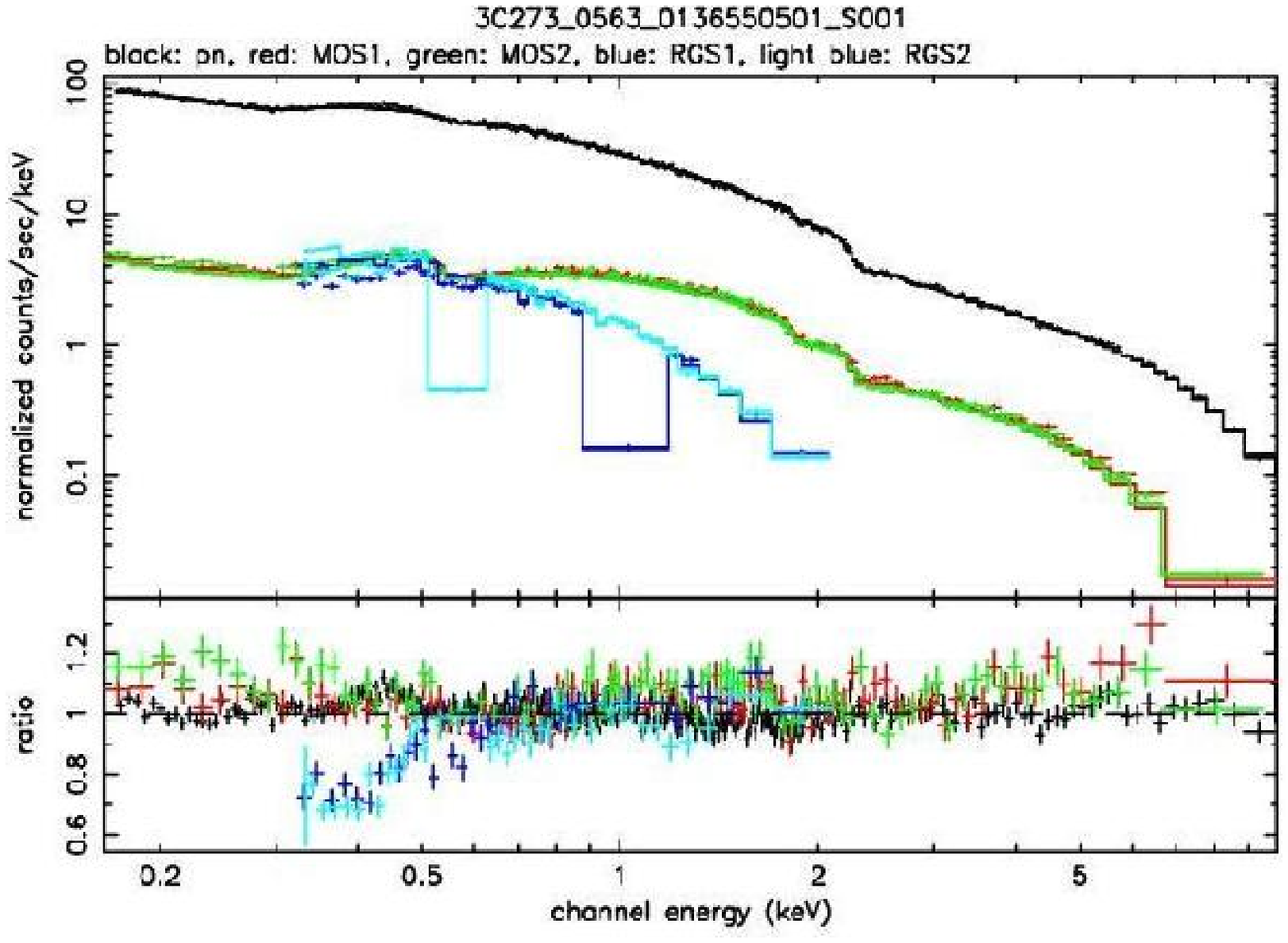}
\includegraphics[width=0.8\linewidth]{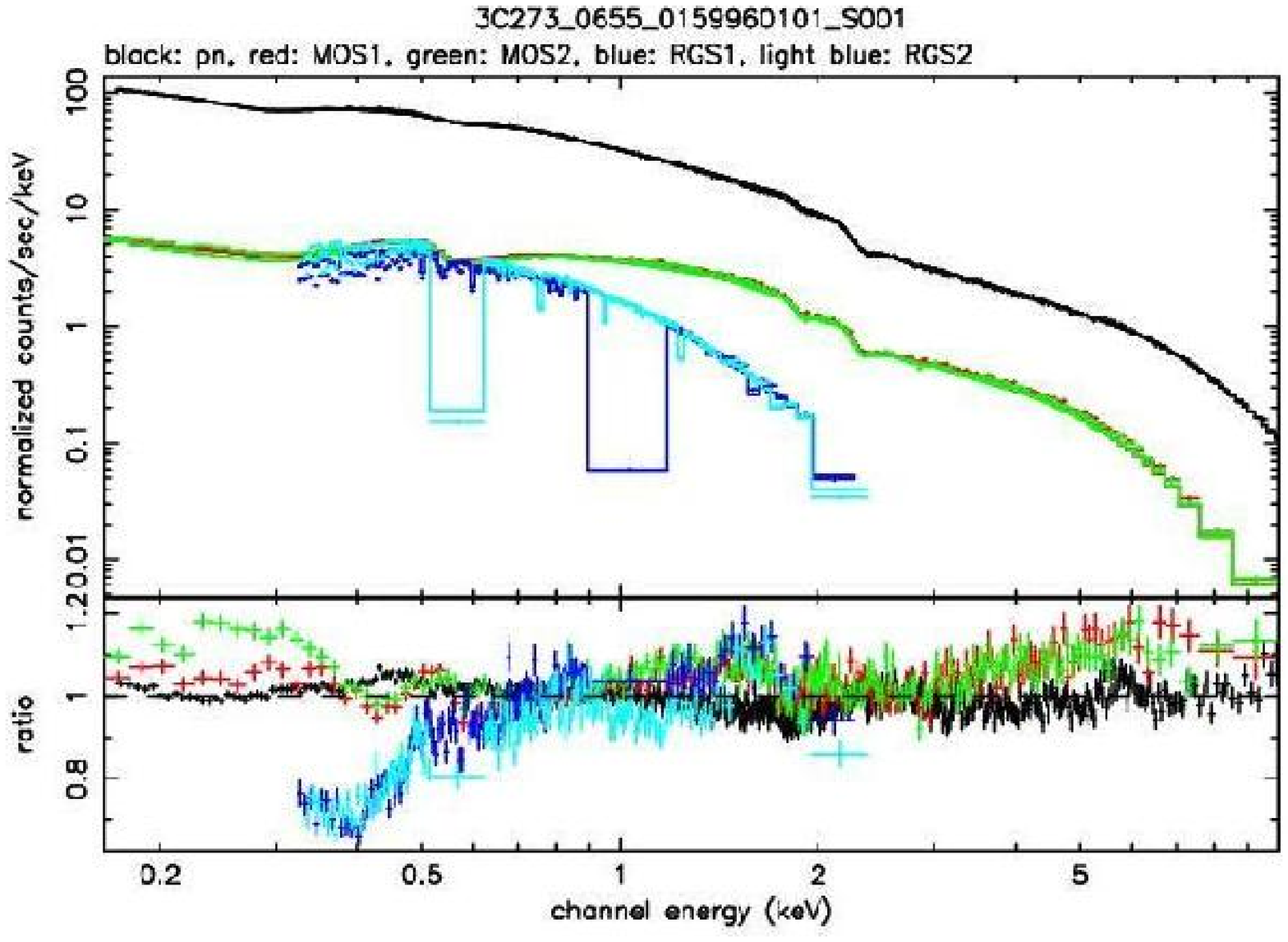}
\includegraphics[width=0.8\linewidth]{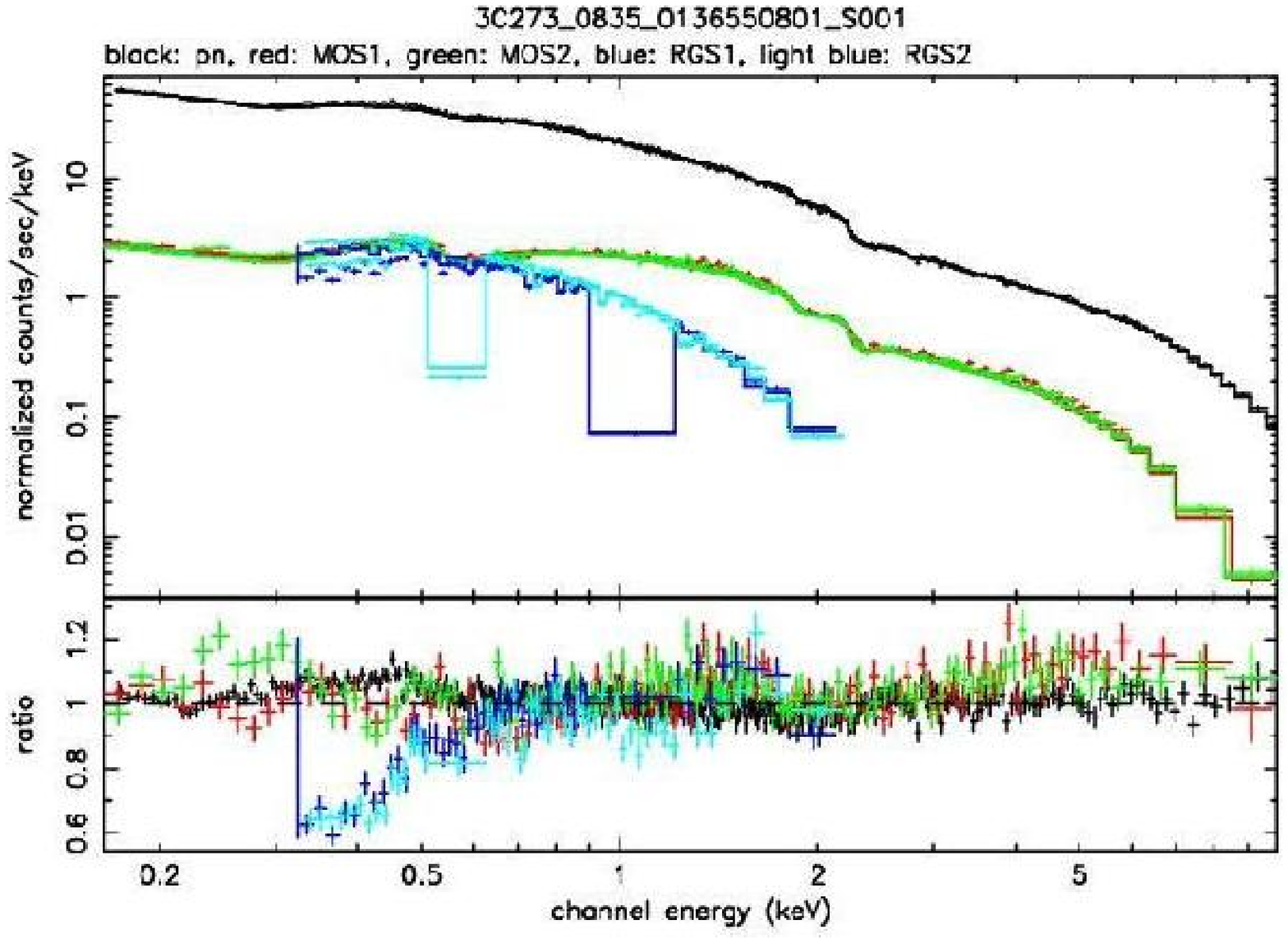}
\includegraphics[width=0.8\linewidth]{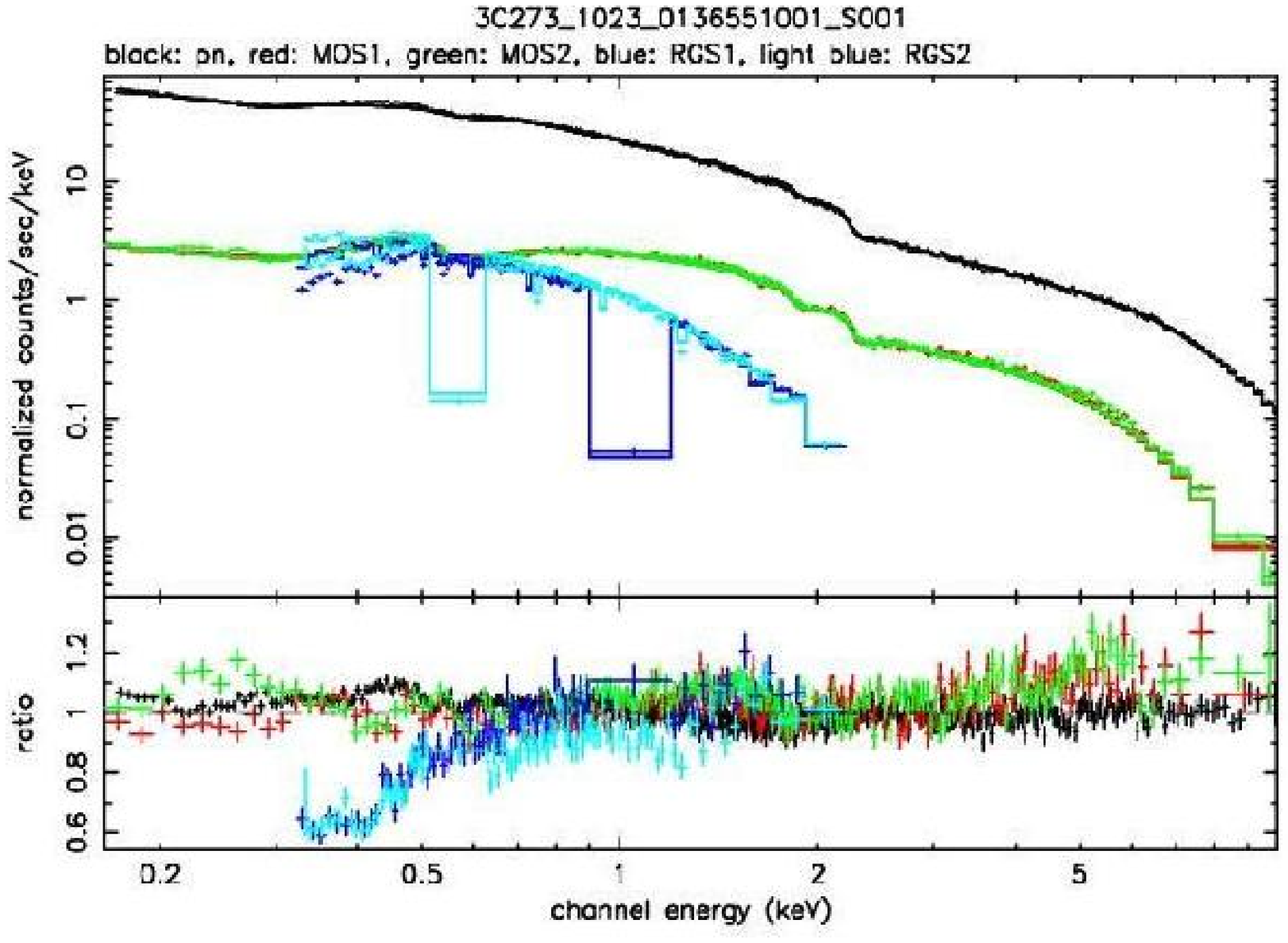}
\caption{Continuum source 3C~273 for the different time epochs
  rev. 277, 563, 655, 835 and 1023 (top to bottom). All observations
  were performed in EPIC small window modes and the medium filters.  
\label{fig:3c273epochs}}
\end{figure}
\end{small}
\begin{small}
\begin{figure}
\centering
\includegraphics[width=0.8\linewidth]{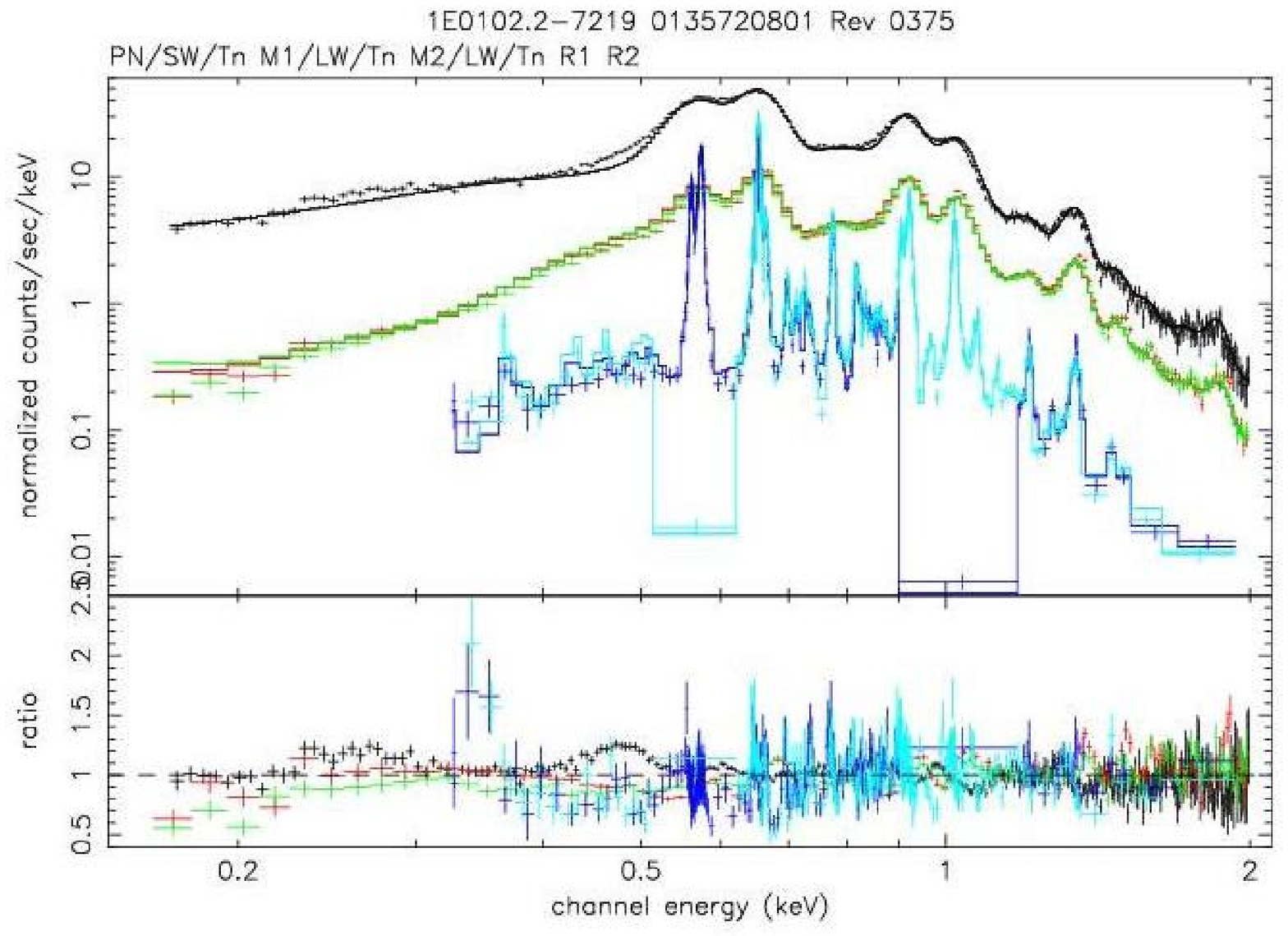}
\includegraphics[width=0.8\linewidth]{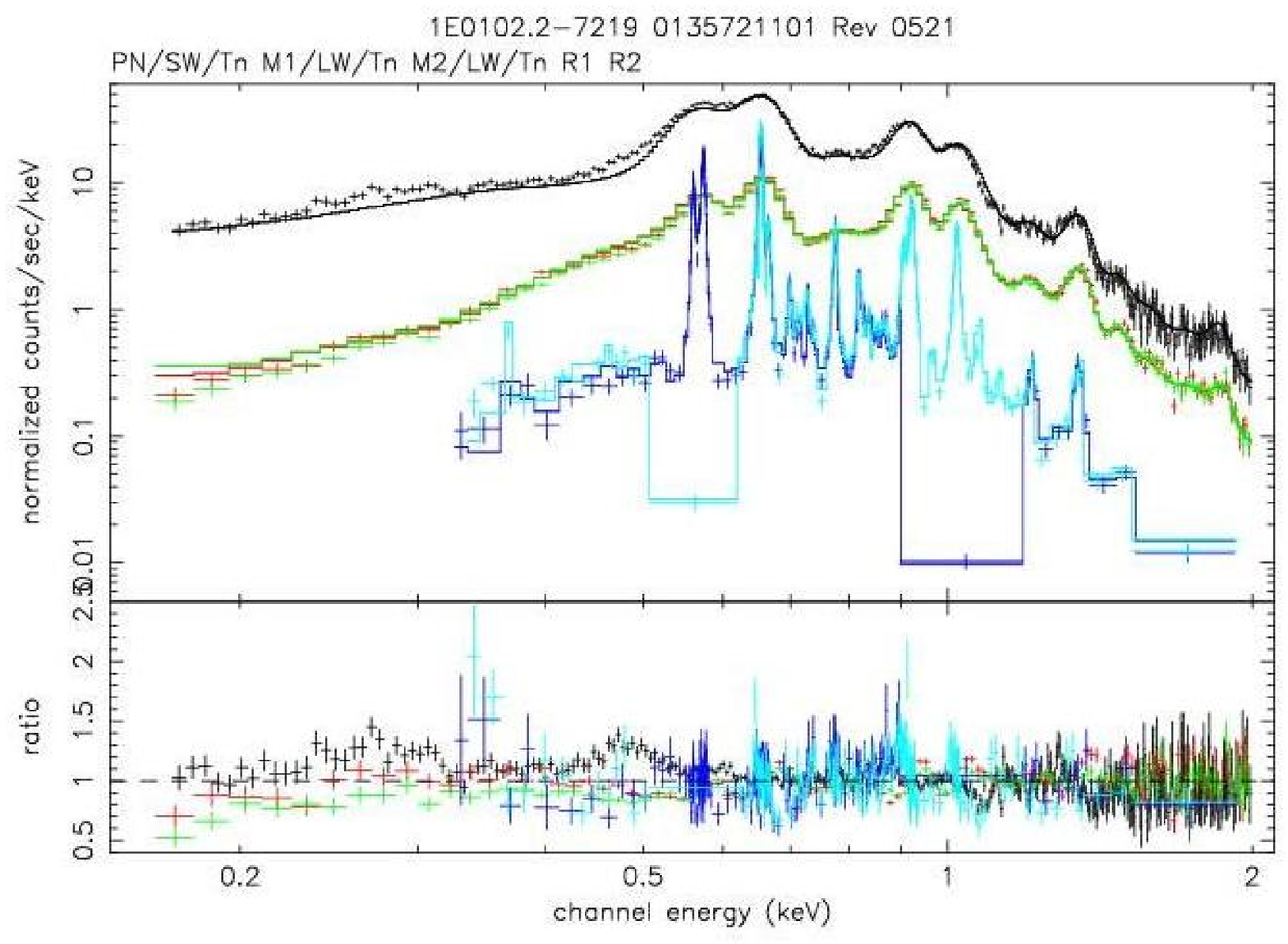}
\includegraphics[width=0.8\linewidth]{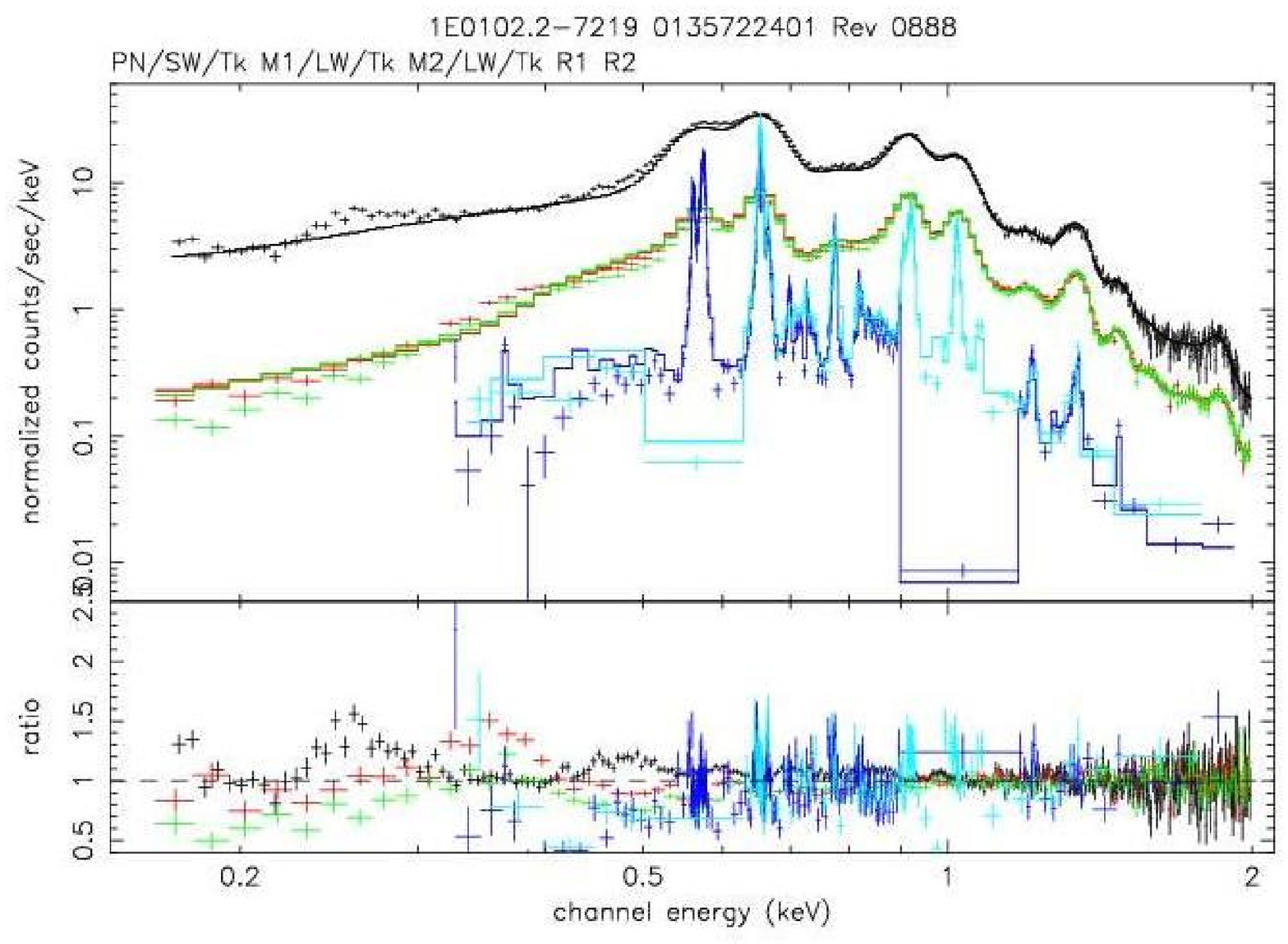}
\includegraphics[width=0.8\linewidth]{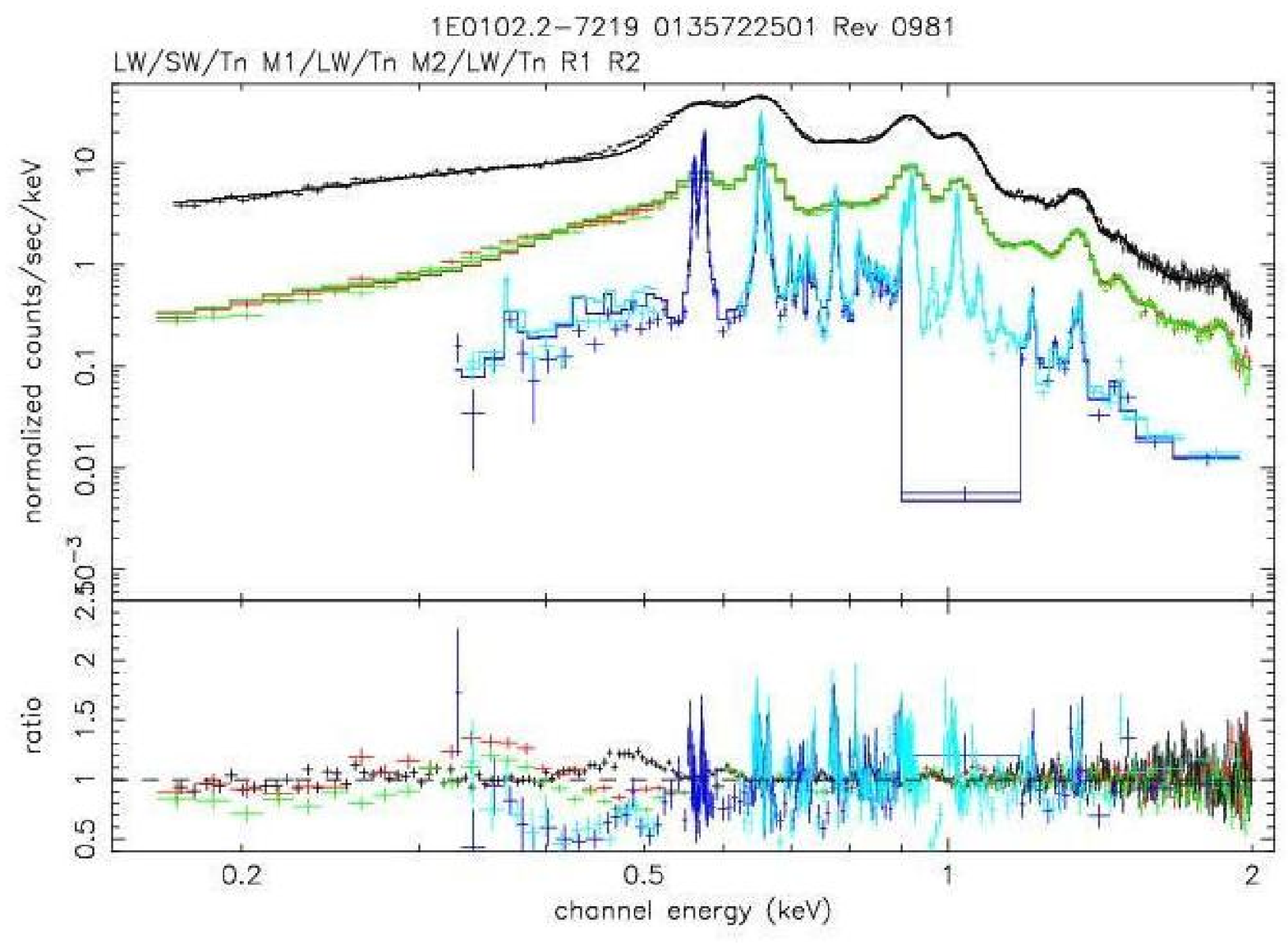}
\caption{Line source 1ES0102-7219 for the different time epochs
rev. 375, 521, 888 and 981 (top to bottom). All observations were
performed in small window mode for pn and the large wondow modes for
MOS. In rev.~888 the thick filters were used, all others were using the
thin filters. 
\label{fig:1es0102epochs}}
\end{figure}
\end{small}

The problem is thought to be not related to the filter model
but to the redistrubution at small count statistics.

\section{Examples of other sources}
In this section we present examples of different sources at different
epochs. PKS0558-504 (Fig.~\ref{fig:pks0558}) was observed in rev.~153,
an epoch before the MOS patch was present. The pn was in small window
mode, MOS in large window mode. All EPICs used the thin filter. This
observation is an example of a perfect agreement between all EPIC
instruments. 

\begin{small}
\begin{figure}[h]
\centering
\includegraphics[width=0.9\linewidth]{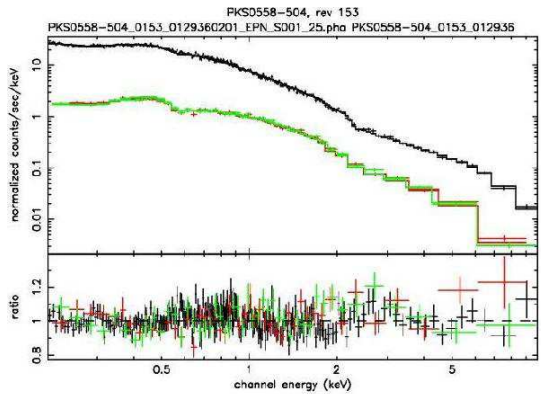}
\caption{PKS0558-504 in rev.~153, before the MOS patch was
  present. Perfect agreement between all EPIC instruments.
\label{fig:pks0558}}
\end{figure}
\end{small}

At the PKS2155-304 observation in rev.~545 (Fig.~\ref{fig:pks2155})
and for H1426+428 in rev.~1012 (Fig.~\ref{fig:h1426}), the MOS patch
was present. Both observations used the EPIC small window modes with
the medium filter. The spectra show a good general agreement for the
total energy range. The largest discrepancies are present between
0.4-0.8~keV, where the MOS are lower than pn by 10-15\%. Above 5~kev,
the pn is lower than the MOS by up to 10\%.

\begin{small}
\begin{figure}[h]
\centering
\includegraphics[width=0.9\linewidth]{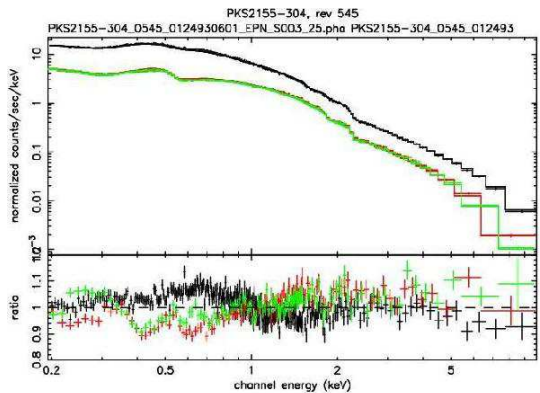}
\caption{PKS2155-304 in rev.~545. with MOS patch present. Largest
  discrepancies present between 0.4-0.8~keV.
\label{fig:pks2155}}
\end{figure}
\end{small}

\begin{small}
\begin{figure}[h]
\centering
\includegraphics[width=0.9\linewidth]{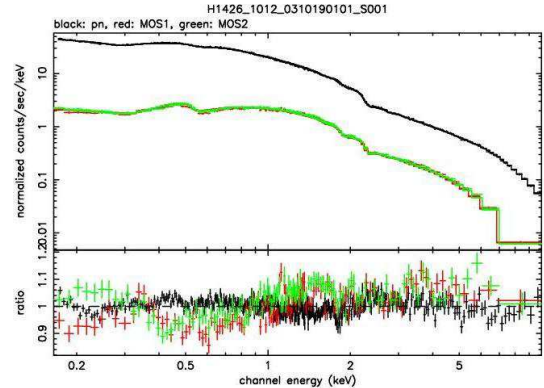}
\caption{H1426+428 in rev.~1012. Discrepancies are reduced to max. 10\%.
\label{fig:h1426}}
\end{figure}
\end{small}

The isolated neutron star RXJ1856-3754 has been observed in rev.~427,
878 and 968 using pn small window mode and the thin filter. The flux
variations were less than 1\%, proving the low energy stability
of the EPIC pn. With SASv6.1, the absorption column was fitted to zero
for pn. Using the new CCFs published 2005 May, the N$_{\rm H}$ is not
disappearing any more, even if the fitted value is lower than the
value obtained from the deep Chandra
observation. Fig.~\ref{fig:rxj1856} presents the rev.~878 observation
with all EPICs in small window mode and thin filters. MOS and pn agree
within 10\% down to lowest energies.
\begin{small}
\begin{figure}[h]
\centering
\includegraphics[width=0.9\linewidth]{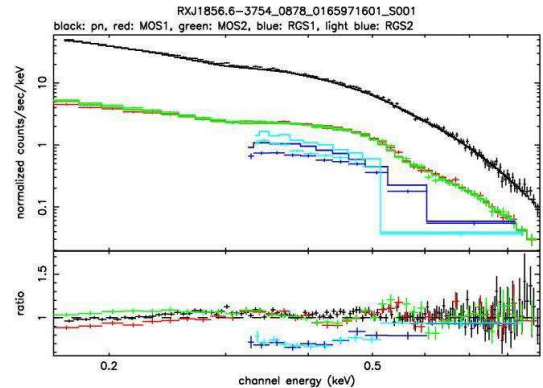}
\caption{RXJ1856-3754 in rev.~878. MOS-pn agreement within 10\% down to
  lowest energies.
\label{fig:rxj1856}}
\end{figure}
\end{small}

\section{XMM-Newton versus Chandra}
The first example presents a simultaneous observation of PKS2155-304
observation in rev.~362 (Fig.~\ref{fig:chandra_letg}). The general
spectral shape measured by XMM-Newton EPIC and
Chandra ACIS/LETG above about 1~keV agrees well, with Chandra
normalisations being higher than the EPIC ones. Below 1~keV and
compared to EPIC, the ACIS/LETG residuals increase to lower energies
whereas the RGS residuals decrease by about the same level.
\begin{small}
\begin{figure}[h]
\centering
\includegraphics[width=0.9\linewidth]{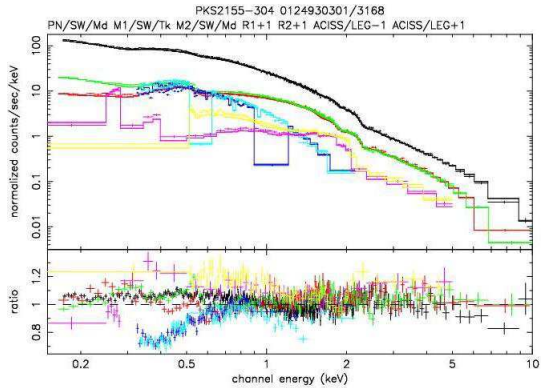}
\caption{Simultaneous XMM-Newton and Chandra observation of
  PKS2155-304 in rev.~362. 
\label{fig:chandra_letg}}
\end{figure}
\end{small}

An example for a simultaneous observation of XMM-Newton with Chandra
ACIS/HEG and ACIS/MEG is presented in
Fig.~\ref{fig:chandra_metg}. H1426+428 was observed in
rev.~1015. Again, the general spectral shape agrees very well between
XMM-Newton and Chandra. Especially between 0.8-2.0~keV  all
instruments agree within 15\% in normalisation. Above 2~keV, the
ACIS/HEG shows a slightly flatter slope than EPIC. At high energies,
MOS are closer to ACIS/MEG than pn.
\begin{small}
\begin{figure}[h]
\centering
\includegraphics[width=0.9\linewidth]{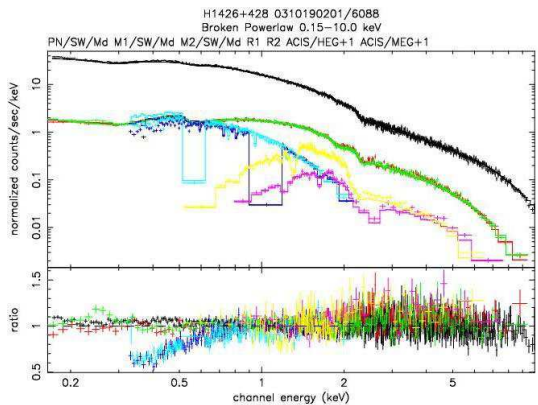}
\caption{Simultaneous XMM-Newton and Chandra observation of H1426+428
  in rev.~1015. 
\label{fig:chandra_metg}}
\end{figure}
\end{small}

\section{XMM-Newton versus Swift}
A simultaneous observation of XMM-Newton and Swift was performed on
H1426+428 in rev.~1012 and the result is presented in
Fig.~\ref{fig:swift}. The residuals show a good agreement of all
instruments between 0.6-3.0~keV. The Swift XRT measures a steeper
slope at high energies than the EPICs. Below 0.6~keV, large
discrepancies are present. 
\begin{small}
\begin{figure}[h]
\centering
\includegraphics[width=0.9\linewidth]{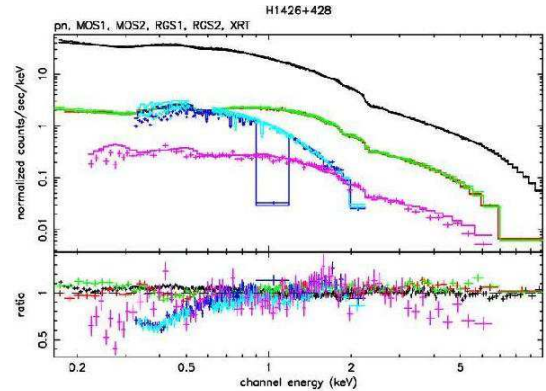}
\caption{Simultaneous XMM-Newton and Swift observation of H1426+428 in
  rev.~1012. Swift data courtesy of Sergio Campana.
\label{fig:swift}}
\end{figure}
\end{small}

\section{Conclusions}
With the new mechanisms for time- and spatial dependent MOS
redistributions in SASv6.5, together with the published corresponding
new set of EPIC MOS-CCFs, the EPIC low energy issue is about to be
solved. The new pn CCFs are already available since 2005 May. RGS show
a low energy flux difference of about 10\% at launch. The sensitivity
at longest wavelengths has decreased by 10-20\% over the mission. The
status using SASv6.5 is that EPIC and RGS are not yet
consistent below 0.7~keV.

The next big step in cross-calibration will be the implementation of
the best knowledge of RGS into SAS (J.~Kaastra et al.:
Absolute Effective Area Calibration of the XMM-Newton Reflection
Grating Spectrometers).

\end{document}